\documentclass{jfm}

\usepackage{hyperref}   % [breaklinks,colorlinks]
\usepackage{graphicx}
\usepackage{epstopdf, epsfig}
\usepackage[french,english]{babel}
\usepackage{float}

\usepackage{enumerate}
\usepackage{amsmath}
\usepackage{amssymb}
\usepackage{floatflt}
\usepackage{float}
\usepackage{fancybox}
\usepackage{color}
\usepackage{sidecap}
\usepackage[font=footnotesize,labelfont=bf]{caption}
\usepackage{subcaption}
\usepackage{siunitx}
\usepackage{caption}

\renewcommand{\cite}{\citep}
\newcommand{\be}{\begin{equation}}
\newcommand{\ee}{\end{equation}}
\newcommand{\bsub}{\begin{subequations}}
\newcommand{\esub}{\end{subequations}}
\newcommand{\bea}{\begin{eqnarray}}
\newcommand{\eea}{\end{eqnarray}}
\renewcommand{\vec}[1]{\mathbf{#1}}
\newcommand{\bnab}{\boldsymbol{\nabla}}

\shorttitle{Waves on a vortex: rays, rings and resonances}
\shortauthor{T. Torres, A. Coutant, S. Dolan and S. Weinfurtner}

\title{Waves on a vortex: rays, rings and resonances}

\author{
Theo Torres
\aff{1}
\aff{2}
\corresp{\email{theo.torresvicente@nottingham.ac.uk}},
Antonin Coutant
\aff{1}
\aff{2}
\corresp{\email{antonin.coutant@nottingham.ac.uk}},
Sam Dolan
\aff{3}
\corresp{\email{s.dolan@sheffield.ac.uk}}
\and 
Silke Weinfurtner
\aff{1}
\aff{2}
\corresp{\email{Silke.Weinfurtner@nottingham.ac.uk}}
}

\affiliation{\aff{1}School of Mathematical Sciences, University of Nottingham, University Park, Nottingham, NG7 2RD, UK
\aff{2}
Centre for the Mathematics and Theoretical Physics of Quantum Non-Equilibrium Systems, University of Nottingham, Nottingham NG7 2RD, UK
\aff{3}Consortium for Fundamental Physics, School of Mathematics and Statistics, University of
Sheffield, Hicks Building, Hounsfield Road, Sheffield S3 7RH, UK
}
\begin{document}

\maketitle

\begin{abstract}
We study the scattering of surface water waves with irrotational draining vortices. At small depth, this system is a mathematical analogue of a rotating black hole and can be used to mimic some of its peculiar phenomenon. Using ray-tracing methods, we exhibit the existence of unstable orbits around vortices at arbitrary depth. These orbits are the analogue of the light rings of a black hole. We show that these orbits come in pairs, one co-rotating and one counter-rotating, at an orbital radius that varies with the frequency. We derived an explicit formula for this radius in the deep water regime. Our method is validated by comparison with recent experimental data from a wavetank experiment. We finally argue that these rings will generate a discrete set of damped resonances that we characterize and that could possibly be observed in future experiments.
\end{abstract}

\begin{keywords}
black hole, surface gravity waves, wave scattering, waves in rotating fluids
\end{keywords}

\section{Introduction}
The interaction between waves and vortices has attracted attention in various fields of physics~\citep{Vivanco00}, such as acoustics~\citep{Fetter64,Nazarenko95,Pagneux01,Kopiev10}, ocean surface waves~\citep{Buhler,Buhler05}, and wave generation by turbulence~\citep{Lund89,Cerda93,Fabrikant94,Umeki97}. The problem was studied with the use of singularities in the complex plane by \cite{Vandenbroeck87,Stokes08,Moreira12}. The stability analysis of rotating fluids is also a rich topic, in particular due to the possibility of ``over-reflection'' mechanisms \citep{Acheson76}, which was studied both experimentally \citep{Vatistas94,Jansson06} and theoretically \citep{Tophoj13,Mougel17}. 

The wave-vortex interaction is the setting for two intriguing physical analogies. Firstly, waves on a vortex give rise to a hydrodynamical analogue of the Aharonov-Bohm effect of quantum mechanics~\citep{Berry_AB,Coste99,Coste99b,Vivanco99,Sonin02,Dolan11}. Secondly, a draining vortex constitutes an (inexact) hydrodynamical analogue of a rotating black hole. Based on the original work of ~\cite{Unruh:1980cg}, \cite{SCH02} showed that (under reasonable physical assumptions) the propagation equation governing waves on a draining vortex is formally identical to that governing a massless scalar field on an `effective' (but fictitious) black hole geometry. Further, the characteristics of the wave equation map on to the null geodesics (i.e.~the light rays) of the black hole geometry. The presence of an horizon -- a one-way membrane for rays and waves -- leads to exotic wave phenomena such as superradiance (a particular case of over-reflection) or Hawking radiation (for reviews see e.g.~\cite{BRI15} on superradiance, \cite{Jacobson12} on Hawking radiation and \cite{Barcelo05} on analogue gravity). Analogous effects are expected in shallow fluid flows when the speed of the flow becomes comparable to the propagation speed of the waves. 

Recently, there has been much interest in studying black hole analogue phenomenology in the laboratory. Amongst others, \cite{Silke_exp}, \cite{Euve15} and \cite{Steinhauer:2015saa} conducted laboratory-based experiments to mimic the analogue Hawking effect of a black hole. An analogue rotating black hole was exhibited in an optical system by \cite{Vocke17}. \cite{Superradiance} have reported the first observation of rotational superradiance, in a draining vortex. Notably, in this last experiment the dispersive nature of the waves needs to be taken into account, and thus the effective-geometry description of \cite{SCH02} is incomplete. Due to a non-linear dispersion relation, the phase and group velocities are \emph{not} equal outside the shallow regime; quite unlike the case in relativity, where both are equal to the speed of light. Nevertheless, the key features predicted for the shallow-water, linear dispersion regime, such as superradiance and the Aharonov-Bohm effect, appear to be robust in real experiments, as observed by \cite{Berry_AB} and \cite{Superradiance} for example. 

An ubiquitous feature of a black hole is the existence of photon orbits called `light rings'. Outside a Schwarzschild black hole, for example, there is a `photon sphere', at $3/2$ times the event horizon radius, comprising the (zero-measure) set of light rays (null geodesics) that orbit around the black hole perpetually. These orbits comprise a separatrix in phase space, dividing between rays that fall into the black hole, and rays that escape. Analysis of the properties of light rings yields a heuristic understanding of many black hole phenomena. The orbital frequency $\Omega$ and Lyapunov exponent $\Lambda$ of the light-rings -- just two parameters -- play a dominant role in (semi-classical approximations to) the spectrum of damped resonances known as black hole quasi-normal modes ~\citep{Goebel72,Cardoso,Dolan_QNM,Yang:2012he,Konoplya17}; the absorption cross section ~\citep{Decanini:2011xi, Macedo:2013afa}; and wave-interference phenomena such as `orbiting' and `glories' in the scattering cross section ~\citep{Matzner:1985rjn, Dolan:2017rtj}. For example, the frequency and decay rate of the `ringdown phase' in gravitational wave chirp signals, such as GW150914 detected by \cite{LIGO16}, are determined by the black hole quasi-normal mode frequencies (\cite{Schutz, Berti:2009kk}).

A key aim of this paper is to show that there generically exist circular orbits around a draining vortex, which are the analogue of black hole light rings. To do so, we apply ray tracing methods, well-used in hydrodynamics, astronomy and elsewhere, to analyse perturbations of real fluid-mechanical systems, such as vortices, in which dispersion and/or dissipation play a significant role. In section~\ref{Method_Sec}, we apply the methods to show that scattering of waves in a draining vortex can be well-understood in terms of the rays of a frequency-dependent effective Hamiltonian, and we obtain a transport equation for the wave amplitude. In section~\ref{LR_Sec} we discuss the presence of circular orbits and in section~\ref{QNM_Sec} discuss its consequences in terms of resonance appearing in the time dependent response of the system. 

\section{Surface waves on a flow}

We consider surface waves propagating on a flow of constant depth $h_0$. 
The flow is assumed incompressible ($\rho = \text{const}$) and irrotational, so that the velocity flow is derived from a potential $\vec{v}=\bnab \Phi$. On top of a background (mean) flow $\vec{v_0}=\bnab \Phi_0$, small perturbations $\phi = \Phi - \Phi_0$ propagate according to the wave equation~\citep{Milewski96}
\be \label{wave_eq}
\mathcal{D}_t^{2}\phi - i (g \bnab  - \gamma \bnab ^3) \tanh(-ih_0 \bnab ) \phi + 2\nu \nabla^2 \mathcal{D}_t \phi = 0,
\ee
where $\mathcal{D}_t=\partial_t + \vec{v_0}.\bnab $ is the material derivative, $g$ is the gravitational acceleration, $\gamma$ is the ratio of surface tension and the fluid density, $\nu$ the viscosity of the fluid, and $i$ the imaginary number such that $i^2 = -1$. In this equation, we included the effect of bulk dissipation, governed by the last term, but other sources of dissipation are generally present~\citep{Hunt64,Przadka12,Robertson17}. To keep the discussion general, we introduce a pair of functions, $F$ and $\Gamma$, encoding the dispersion relation and dissipation, in order to write the wave equation in the following form: 
\be
\mathcal{D}_t^{2}\phi +F(-i\bnab ) \phi - 2 \nu \Gamma(-i\bnab ) \mathcal{D}_t \phi = 0.
\ee
In the absence of background flow, solutions of \eqref{wave_eq} are given by plane waves \mbox{$\phi = \Re [A \exp(-i \omega t + i \vec k \cdot \vec x)]$} of amplitude $A$, angular frequency $\omega$ and wave vector $\vec k$ (of norm $k = ||\vec k||$). $\omega$ and $\vec k$ then obey the dispersion relation
\be
\omega^2 = F(\vec k) - 2 i \nu \omega \Gamma(\vec k), 
\ee
with the functions 
\be  \label{dispersion_relation}
F(\vec k) = (g k + \gamma k^3) \tanh(h_0 k) \qquad \textrm{and} \qquad \Gamma(k)=k^2.
\ee
Note that the method we shall present stays identical for general functions $F$ and $\Gamma$, the only necessary assumptions are that they are analytic and even. In the absence of background flow, a wave packet propagates at a group velocity $\vec{v}_{g}$ given by 
\be
\vec{v}_{g} \doteq \bnab_k \omega = \bnab _k \sqrt{F}. 
\ee
In the presence of a constant flow, one must distinguish between two notions of (angular) frequency: the intrinsic (fluid frame) frequency $\Omega$ and the laboratory frequency $\omega$. The intrinsic frequency is the one satisfying the dispersion relation, while the laboratory one is conserved due to stationarity. The two are related by a Doppler shift: $\Omega=\omega - \vec{v_0} \cdot \vec{k}$. As a result, the group velocity in the lab frame is defined as
\be
\vec{v}_{g}^{\rm lab} = \bnab_k \omega,
\ee
and obtained from the group velocity in the fluid frame using 
\be
\vec{v}_g^{f} \doteq \bnab_k \sqrt{F} = \vec{v}_g^{\rm lab} - \vec{v_0}.
\ee
Now when the background flow varies spatially, solutions of the wave equation become much more involved. This is why we shall now solve Eq.~\eqref{wave_eq} using a ray tracing method, to characterize the propagation of waves around vortices.

\section{Ray tracing equations}
\label{Method_Sec}

We shall now solve Eq.~\eqref{wave_eq} using a gradient expansion (a common method in wave physics, see e.g.~\cite{Buhler}). This means that we assume that the background flow changes over a scale significantly larger than the wavelength. To materialize this, a convenient procedure is to rescale the derivatives in \eqref{wave_eq} as $\partial \to \epsilon \partial$, and solve the equation perturbatively in $\epsilon$. We then look for wave solutions with a rapidly varying phase and a slowly varying amplitude, that is, we seek solutions of the form  
\be \label{eikonal_wave}
\phi = A(\vec x) \exp \left(i\frac{S(\vec x)} \epsilon\right). 
\ee
$A$ is the local amplitude, and $S$ the local phase, both real-valued functions (of course ultimately, one should take the real part of \eqref{eikonal_wave} to obtain the velocity potential fluctuation field). We now expand $S$ and $A$ in powers of $\epsilon$, and obtain an equation for the phase and one for the amplitude (in Appendix~\ref{WKB_App}, we detail how to obtain them). At leading order in $\epsilon$, the phase $S_0$ obeys the Hamilton-Jacobi equation, 
\begin{equation} \label{H-J}
\left(\partial_{t}S_{0} + \vec{v_0} \cdot \bnab S_{0}\right)^{2} - F(\bnab S_{0}) = 0.
\end{equation}
This first order partial differential equation can be solved by the method of (bi)characteristics. The characteristics are the {\it rays} of the waves. 
Substituting the $\bnab S_0$ by the wave vector $\vec k$ and $\p_t S_0$ by $-\omega$, we obtain the Hamiltonian of our system, 
\begin{equation} \label{disp_Ham}
\mathcal{H} = -\frac{1}{2}\left( \omega - \vec{v_0} \cdot \vec{k} \right)^{2} + \frac{1}{2} F(\vec k).
\end{equation}
The characteristics are now obtained through Hamilton's equations:
\bsub
\label{Ham_eq}
\bea
\dot t &=& -\frac{\partial \mathcal{H}}{\partial \omega}, \qquad \dot \omega =  \frac{\partial \mathcal{H}}{\partial t} , \label{Ham_eq_om}\\
\dot{x_j} &=& \frac{\partial\mathcal{H}}{\partial k_j} \qquad \text{and} \qquad \dot{k_j}=-\frac{\partial \mathcal{H}}{\partial x_j}, \label{Ham_eq_x}
\eea
\esub
where the dot denotes the derivative with respect to a parameter $\lambda$, used to parametrize the rays, and the index $j$ is short for $(x,y)$. (The unusual sign in \eqref{Ham_eq_om} is due to the fact that the conjugate variable to $t$ is $-\omega$ and not $\omega$.) In addition to Hamilton's equations, one must ensure that the Hamiltonian constraint 
\be
\mathcal{H} = 0, 
\ee
is satisfied for all solutions. A satisfying feature of this constrained Hamiltonian formalism is that it is independent of the choice of parameter. One can multiply the Hamiltonian $\mathcal H$ by any non-vanishing function $N(\lambda)$, and one will the find the same set of rays, instead parametrized by $\tilde \lambda$ where
\be
d\tilde \lambda = N(\lambda) d \lambda. 
\ee 
The next-to-leading order expansion of Eq.~\ref{wave_eq} in $\epsilon$  gives us the equation for the amplitude. This takes the form of a transport equation, 
\be \label{transport_eq}
\partial_t \left( \Omega_0 A_0^2 \right) + \bnab \cdot ( \Omega_0 A_{0}^{2} \vec{v}_g^{\rm lab}) = - 2 \nu \Omega_0 A_0^2  \gamma(\bnab S_0) 
\ee
Note that this equation can be solved once one has a solution $S_0$ of the Hamilton-Jacobi equation \eqref{H-J}. In particular, the group velocity in Eq.~\eqref{transport_eq} is only a function of the position: $\vec{v}_g (\vec x,p) \rightarrow \vec{v}_g(\vec x,\nabla S_0 (\vec x))$. The quantity $A_0^2 \Omega_0$ is in fact the wave action \citep{Buhler}, up to a factor of $\rho / g$. This equation implies that the wave action is moving with the velocity $\vec{v}_g^{\rm lab}$. The dissipative term simply encodes the loss of wave action carried along the propagation. Since dissipation does not affect the equations \eqref{Ham_eq} that determine the paths of the rays, we neglect it in the following.

Notice that we have so far considered a constant surface height. However, it is formally simple to add a varying height $h_0(x,y)$ at the level of Hamilton's equations \eqref{Ham_eq}. Such replacement will be valid if the variations of the surface height are slow enough. By this we mean that first, the relative variations of the height are small $|\bnab h_0| \ll 1$, so that its gradient can be neglected in the derivation of the wave equation \eqref{wave_eq}. And second, that the wavelength is small compared to the variation scale $|\bnab h_0/(k h_0)|$, so that the eikonal approximation stays valid. In the experiment of section \ref{Exp_Sec} the assumption of constant height is well satisfied outside a small radius near the center, and hence we shall assume it in the rest of the paper.

\section{Rays around a vortex}
\label{LR_Sec}
We shall now apply the ray-tracing equations to study the propagation of waves around a vortex. 
For this we consider the simplest model of an irrotational vortex with a drain.
\begin{figure}
\centering
\includegraphics[trim=0.7cm 0 0 0]{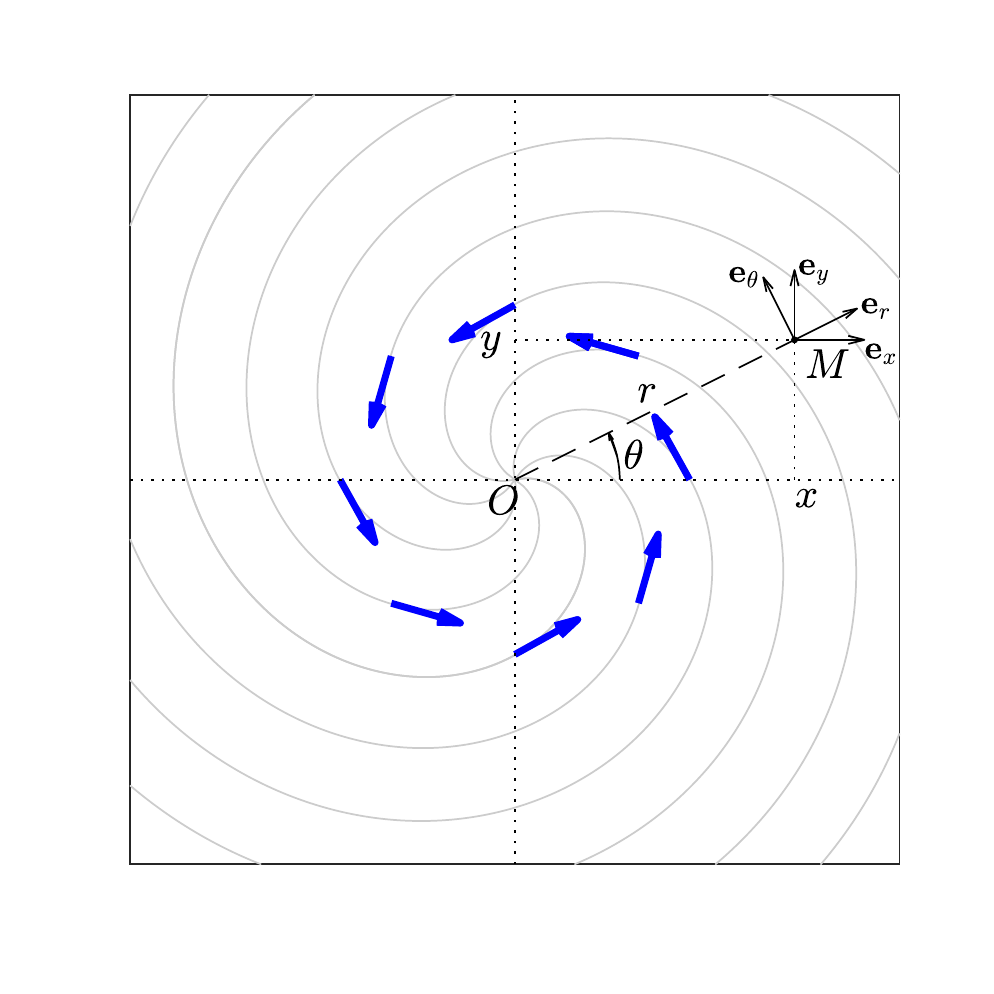}
\caption{Schematic of the flow and associated coordinate systems, from the perspective of an observer looking down on the wavetank. The vortex considered here is rotating counter-clock wise when $C>0$ (as shown by the bold blue arrows). The light gray lines represent streamlines of the flow. A point $M$ on the surface is located using either Cartesian coordinates $(x,y)$ or polar coordinates $(r,\theta)$.
}
\label{schematic}
\end{figure}
This model assumes that the surface is flat, and that the flow is homogeneous in the vertical direction and symmetric around the vertical axis. Under these assumptions, the irrotational and incompressibility conditions gives us the profile 
\be \label{DBT}
\vec v_0 = - \frac Dr \vec e_r + \frac Cr \vec e_\theta, 
\ee
written in polar coordinates $(r,\theta)$ (see Fig.\ref{schematic}), and where $C$ and $D$ are the circulation and draining constants (positive), respectively.
The schematic of the flow and the coordinate systems are presented in Fig. \ref{schematic}. 
This model is not only simple to analyze, it also provides a realistic description of a wide variety of situations. The reason is that far from the center (where the water depth drops), and far from the edges of the fluid tank (where circular symmetry breaks down), the free surface tends to be flat and the flow to ressemble that of Eq.~\eqref{DBT}. As we will see, the effect we are after is contained in this region (see also figure 4 in~\cite{Superradiance}).
This model also corresponds (in the shallow water regime) to a water-wave analogue of a rotating black hole \citep{Visser:1997ux, SCH02}. 

\subsection{Circular orbits\label{subsec:circorbits}}
Although we shall have in mind the flow profile of Eq.~\eqref{DBT}, our results stay conceptually identical for a large class of flows. To see this, the following discussion is done using a general symmetric flow of the form  
\be
\vec v_0 = v_r(r) \vec e_r + v_\theta(r) \vec e_\theta. 
\ee
Using the symmetry of the vortex flow, we first switch to polar coordinates. We build the conjugate variables as before, that is, $k_r = \partial_r S$, and $m = \partial_\theta S$. As we shall see, depending on the context, $m$ will be a real number (interpreted as an angular momentum), or an integer (interpreted as the azimuthal number of the wave). We also note that $\omega$ is conserved since the flow is stationary. For simplicity, we assume $\omega > 0$ (but negative values can be obtained via the symmetry $\omega \to - \omega$ and $m \to -m$).

A circular orbit is an equilibrium point in the radial direction. This means that it is a critical point of the Hamiltonian for $(r,k_r)$. It thus satisfies the conditions
\bsub \label{CO_eq} \bea 
\partial_r \mathcal H &=& 0, \\
\partial_{k_r} \mathcal H &=& 0,
\eea \esub
In the bathtub profile of Eq.~\eqref{DBT}, Eqs.~\eqref{CO_eq} give 2 relations between 4 unknowns. For a fixed pair $(\omega , m)$, Eqs.~\eqref{CO_eq} determines a unique pair $(r_\star, k_{r \star})$. Furthermore, imposing the Hamiltonian constraint $\mathcal H = 0$ gives a relation between $\omega$ and $m$ on the circular orbit, analogous to a dispersion relation, which we write as
\be \label{LR_disp}
\omega = \omega_{\star}(m). 
\ee
This relation can be read in either ways. At fixed angular frequency $\omega$, it gives the value of $m$ such that one has a circular orbit; at fixed azimuthal number $m$ it yields the particular wave frequency corresponding to the circular orbit.

\subsubsection{Shallow water regime}
\label{LR_shallow_Sec}
In the limit of shallow water, where all modes propagate at the same speed, the  radius of the circular orbit is \emph{independent} of the angular frequency $\omega$, and furthermore the relationship between $\omega$ and $m$ is linear (see e.g.~\cite{Dolan12} or \cite{Dempsey2017}): 
\be \label{om-shallow}
\omega_\star = \hat{\Omega}^{\pm} m .
\ee
Here $\hat{\Omega}^\pm \equiv \dot{\theta} / \dot{t}$ is the orbital frequency of the circular orbit. In general, there exist two circular orbits, one co-rotating ($\dot{\theta} > 0$, $\hat{\Omega}^+ > 0$) and one counter-rotating  ($\dot{\theta} < 0$, $\hat{\Omega}^- < 0$) relative to the circulating flow. The frequencies are given by
\be
\hat \Omega^{\pm} = \pm \frac{c^2 \sqrt{C^2+D^2}}{B_{\pm}^2} . 
\ee
The radii of these circular orbits (that we shall now refer to as `orbital radii') are then given by 
\be \label{radius_rel}
r_\star^{\pm} = \frac{B_{\pm}}{c},
\ee
where $c = \sqrt{gh}$ is the propagation speed of shallow water waves, and a $+$ ($-$) sign denotes the co-rotating (counter-rotating) case. The parameter $B_{\pm}$ is defined as
\be \label{B_param}
B_{\pm} = \left(2(C^2 + D^2) \mp 2C\sqrt{C^2 + D^2} \right)^{1/2}. 
\ee 
As we shall see, $B_\pm$ also governs the orbital radius for deep water waves. An immediate conclusion of \eqref{radius_rel} is that the co-rotating circular orbit is closer to the centre of the vortex, while the counter-rotating orbit is further out. Hence, the counter-rotating orbit is in general more visible (this is further confirmed in section~\ref{QNM_Sec} when studying damped resonances).

\subsubsection{Deep water regime}
For further insight into circular orbits in the presence of dispersion, it is instructive to look at another simple case where analytic formulae can be obtained. Here we consider deep water without capillarity, and thus a group velocity given by
\be 
v_g(k) = \frac12 \sqrt{\frac gk}. 
\ee
In this regime it is possible to derive a simple formula for the radius of the orbits as a function of the frequency.
We can directly express the norm of the wave vector in terms of the radius as
\be \label{p_deep}
|\vec{k}| = \frac{g r_{\star}^2}{4B_\pm^2}.
\ee
(with $\pm$ indices omitted on $k$ and $r_\star$ for clarity). 
With the other conditions for the circular orbits, Eq.~(\ref{CO_eq}), we can expression both $k_{r \star}$ and $m$ as functions of $r_{\star}$, viz.,
\be \label{pr_m_deep}
k_{r \star} =\frac{D g r_{\star}^2}{4 B_\pm^3} \quad \text{and} \quad m = \frac{g r_{\star}^3}{4 B_\pm^2}\sqrt{ 1 - \frac{D^2}{B_\pm^2}}.
\ee
Combining these expressions and using the Hamiltonian constraint, we obtain the radii of the unstable orbits as a function of the frequency and the effective dispersion relation (see appendix \ref{App:orbits} for detail):
\be \label{rad_deep}
r_\star = \frac{8B_\pm}{3g} \omega_\star ,
\ee
and
\be \label{effective_dr}
\omega_\star(m)=\frac{3}{8}\left( \frac{4g^2}{\sqrt{B_\pm^2-D^2}} \right)^{1/3} m^{1/3}.
\ee
Note that the wave angular frequency $\omega_\star$ scales linearly with $m$ in shallow water, but with $m^{1/3}$ in deep water. This is a direct consequence of the dispersive nature of the waves. 

\subsubsection{General case}
For any non-linear dispersion relation, \eqref{CO_eq} implies the relation  
\be \label{Implicit_rad_circ}
r_\star = \frac{B_{\pm}}{v_g^f}. 
\ee
This equation shows that it is the change in the group velocity $v_g^f$ that shifts the location of the circular orbit.
Here $v_g^f$ is evaluated with the local wave vector on the orbit $k^{2} = k_{r \star}^2 + m^2/r_\star^2$, and hence the equation above is implicit. To find $r_\star$ and $\omega_\star(m)$ explicitly, one must insert Eq.~\eqref{Implicit_rad_circ} into Eqs.~(\ref{CO_eq}) and the Hamiltonian constraint. 

\begin{figure}
\centering
\includegraphics{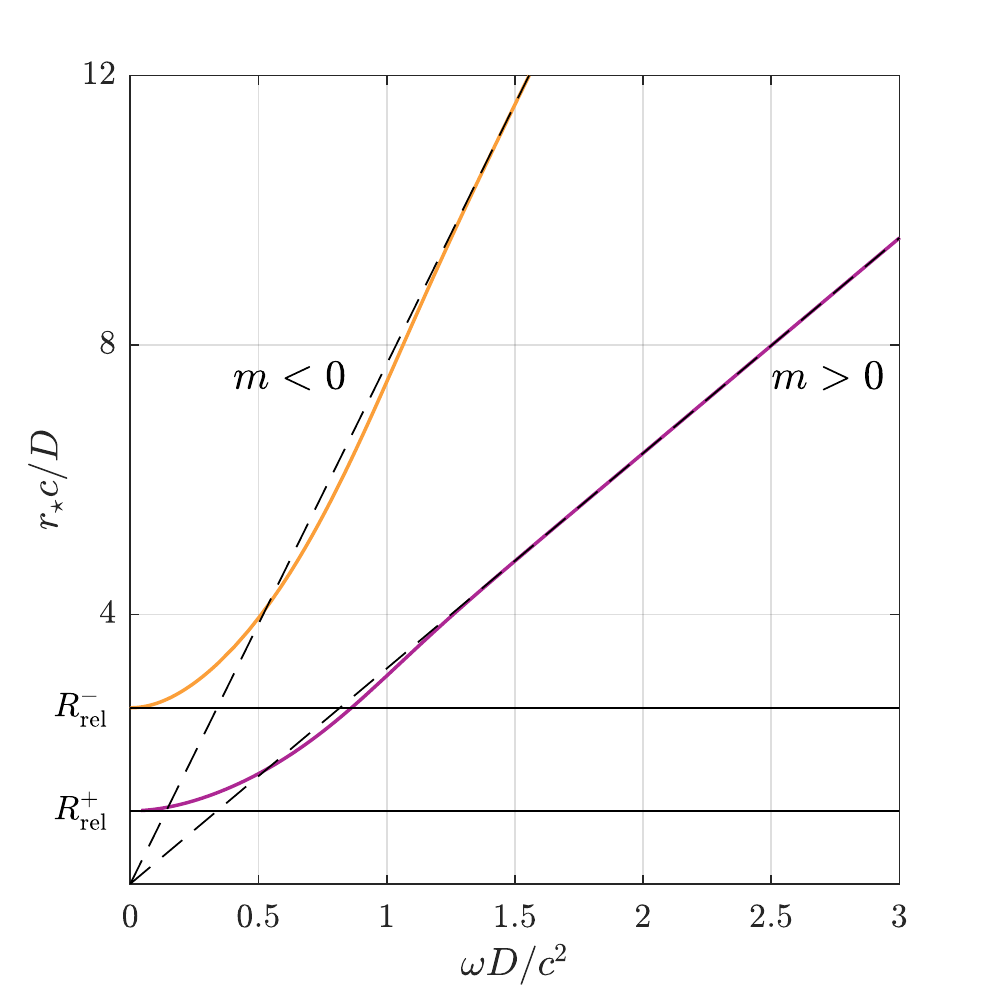}
\caption{Radius of the unstable orbit as a function of the angular frequency of the co- and counter-rotating rays (in purple and orange respectively) and comparison with the shallow and deep water regime (in solid and dashed black respectively). We have chosen the flow parameters such that $C/hc = D/hc = 0.9$ (chosen to make the transition from shallow to deep water more visible), where $c = \sqrt{gh}$ the celerity of shallow water waves.  $R_{\rm rel}^\pm$ is the value of $r_\star^\pm$ in the shallow water (or relativistic) regime. (We recall that when the frequency varies, the corresponding orbital $m$ also does so that \eqref{LR_disp} is satisfied.)
}
\label{radius_orbit}
\end{figure}

Figure \ref{radius_orbit} shows numerical-obtained solutions for the orbital radius as a function of wave frequency, for a vortex with $C/hc = D/hc = 0.9$. The two branches correspond to the co-rotating ($+$) and counter-rotating ($-$) orbits. At low frequency (long wavelength), the orbital radius is anticipated by the shallow-water result, Eq.~(\ref{om-shallow}). At higher frequencies, the circular orbits migrate away from the vortex core, as anticipated in Eq.~(\ref{rad_deep}).
Even though the ray structure will qualitatively be similar for various values of the parameters C and D, one can distinguish two interesting limits. First, when $C \ \gg D$, we can see that the radius of the co-rotating orbit shrink to zeros as $B_{+} \to 0$. At the opposite, when $C \ll D$, the two circular orbits asymptotically approach one another as $B_{+} \to B_{-}$.
 
At this level, it is instructive to discuss the case of a vortex without a drain. By taking the limit $D \to 0$ in equations \eqref{radius_rel} and \eqref{rad_deep}, we see that the co-rotating orbit degenerates to $r=0$, but the counter-rotating one still exists at a nonzero radius (e.g. $C/2c$ in the shallow water regime). 
For sound waves, it was shown by \cite{Nazarenko94} and \cite{Nazarenko95} that rays approaching $r=0$ of an ideal non-draining vortex (equation \eqref{DBT} with $D=0$) see their wavelength decrease to zero, and ultimately reach the dissipative scale.
As such rays do not escape, it suggests that the scattering will be similar to a vortex with small drain (that is $D \neq 0$ but $D \ll C$). In particular, we expect to have a counter-rotating orbit, as well as the associated resonance effects (see section \ref{QNM_Sec})\footnote{We thank an anonymous referee for having stimulated this discussion, and for pointing out references \cite{Nazarenko94} and \cite{Nazarenko95}.}. 

\subsection{Ray scattering}
We now study a family of characteristics (rays) encroaching upon the vortex from the right ($x \to +\infty$) with a angular frequency $\omega$ and various impact parameters $b$, which is the ray equivalent of sending a plane wave. Here, $m$ is a continuous parameter which is given in terms of the impact parameter via the relation 
\be \label{impact_param}
b = \frac{m}{k_{\rm in}} + \frac{C}{v_g^{\rm in}}. 
\ee
Here $k_{\rm in}$ is the \textit{incoming} wavenumber satisfying the dispersion relation (or the Hamiltonian constraint) at infinity for a given $\omega$ and $v_{g}^{\rm in}$ is the group velocity at infinity (as the flow is negligible at infinity, the group velocity in the fluid frame and in the laboratory frame are identical). We numerically solve Hamilton's equations \eqref{Ham_eq} to find the characteristics. A standard fourth-order Runge-Kunta (RK4) scheme is used. To ensure the validity of our numerical simulation, the step size is chosen such that along the rays the adimensional quantity $|\mathcal{H}/ \omega^2|$ is smaller than $ 10^{-7} $. Our solution therefore satisfies the Hamiltonian constraint. 
To numerically compute the amplitude of the wave, we use Eq.~\eqref{transport_eq} with $\nu = 0$. This shows that the flux of wave action is conserved along a tube of rays. Numerically, we use this conservation to obtain the change in amplitude between neighbouring points along two neighbouring rays, by ensuring that the product of the distance between the rays and the wave action is constant. This means that if the rays converge (diverge), the amplitude will increase (decrease). 

\begin{figure}
\begin{subfigure}{.5\textwidth}
\centering
  \includegraphics[width=1\linewidth]{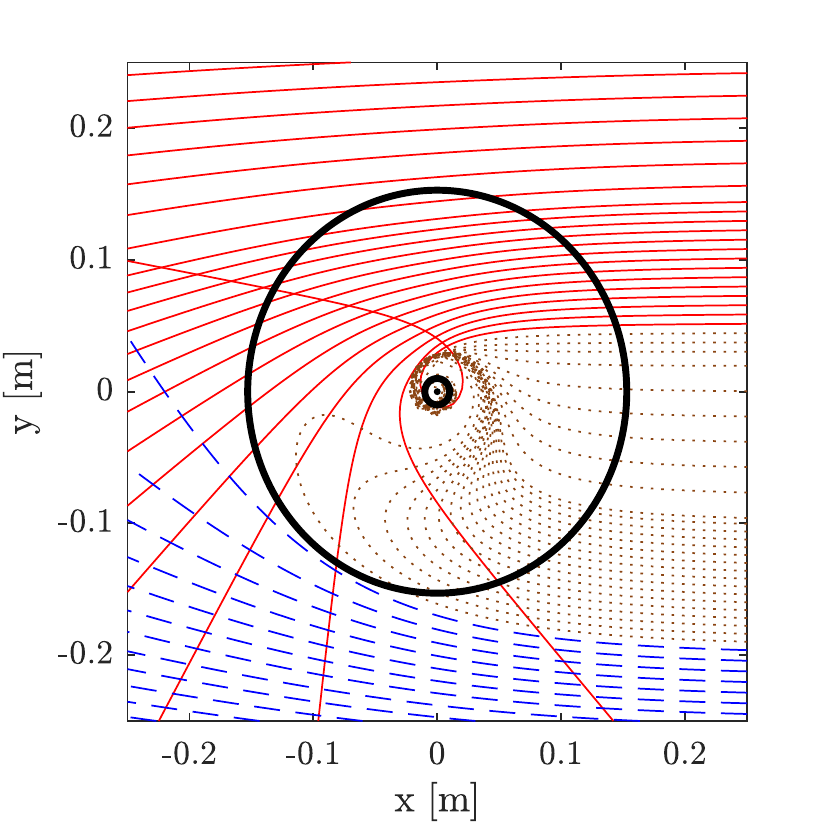}
  \caption{Congruence of rays}
  \label{characteristics}
\end{subfigure}%
\begin{subfigure}{.5\textwidth}
  \includegraphics[width=1\linewidth]{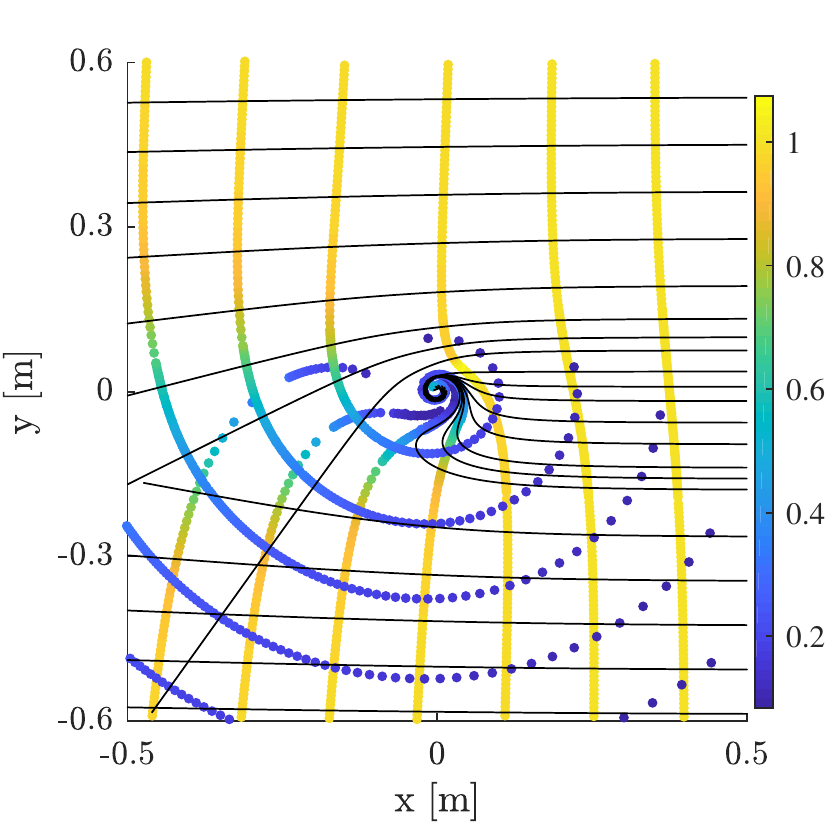}
  \caption{Reconstructed eikonal wavefront}
  \label{wavefront}
\end{subfigure}
\caption{Figure (a) depicts a congruence of rays incoming from far right on a draining vortex flow. The flow parameters are $C=1.6\times10^{-2} \text{m}^{2} \text{s}^{-1}$, $D=1\times10^{-3} \text{m}^{2} \text{s}^{-1}$ and $h=0.06 \text{m}$. They correspond to the experimental flow of \cite{Superradiance}. The angular frequency of the wave is $\omega =19.8 \; \mathrm{rad/s}$. For these parameters, we have $h k_{\rm in} \simeq 2.4$ and hence are in the deep water regime. The red (and blue) curves represent the co-rotating (and counter-rotating) rays escaping to infinity. The dotted brown lines are the rays falling in the hole. The two black circles represent the orbital radius for co- and counter- rotating rays, respectively at $r=1$ cm and $r=15.3$ cm. Figure (b) represents the eikonal wavefront reconstructed from the phase along the rays. The color scale represents the amplitude of the wave normalized to one initially. The black line are some rays from (\ref{characteristics}).}
\label{Rays}
\end{figure}

Figure \ref{characteristics} depicts a congruence of rays at fixed angular frequency $\omega = 19.8 \; \mathrm{rad/s}$ impinging on a draining bathtub vortex with parameters $C=1.6\times10^{-2} \text{m}^{2} \text{s}^{-1}$, $D=1\times10^{-3} \text{m}^{2} \text{s}^{-1}$ and $h=0.06 \text{m}$, corresponding to the laboratory study of \cite{Superradiance}. We can clearly distinguish two types of rays. The first type are rays that are able to escape to infinity (co-rotating in red and counter-rotating in dashed blue). The impact parameter for those rays are beyond some critical values $b_{\rm c}^{\pm}(\omega)$. We note that $|b_{\rm c}^{-}(\omega)|>|b_{\rm c}^{+}(\omega)|$, implying that co-rotating rays are able to go closer to the vortex than counter-rotating ones. The second type of rays are those that fall in to the vortex core (in dotted brown). 
We note from \eqref{impact_param} that the impact parameter depends on the group velocity and therefore on the frequency of the wave. This behaviour differs from the shallow water regime where the group velocity matches the phase velocity and is constant for all frequency.

In the process of obtaining the rays by solving Hamilton's equations, we also compute the phase of the wave along each trajectory. It is then possible to reconstruct the eikonal wavefronts by finding the constant phase points along each ray (see \cite{Dolan_Dempsey} for a shallow-water study). The wavefronts are presented in Fig.~\ref{wavefront}. The coloured dots represent the location of the constant phase points and the colour scale gives the amplitude of the wave at these points. Far from the vortex, where the flow becomes negligible, we can see that the wavefronts are orthogonal to the rays (black lines). On the contrary, closer to the vortex, we clearly observe the non-orthogonality of the wavefronts and the rays where the flow becomes more rapid.

When varying the angular frequency $\omega$, the rays display a similar pattern as in Fig.~\ref{characteristics}, but the orbital radii $r_\star^\pm$ vary. When increasing the angular frequency, they interpolate between their shallow water value to the deep water behaviour of equation~\eqref{rad_deep}. To illustrate this, we plotted the dependence of the orbital radii with the frequency in Fig.~\ref{radius_orbit}.

\subsubsection{Experimental validation}
\label{Exp_Sec}
\begin{figure}
\centering
\includegraphics[trim=0.7cm 0 0 0]{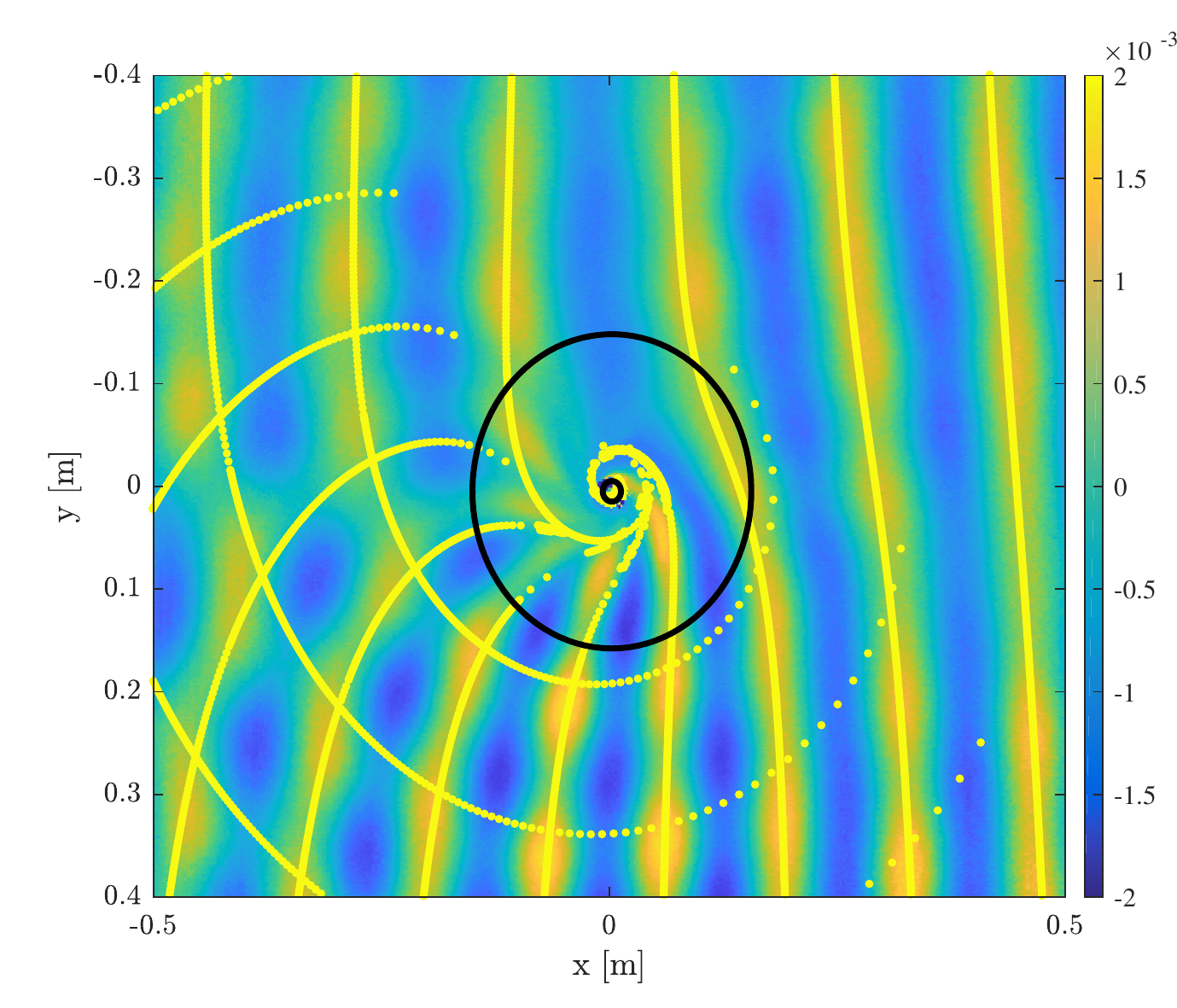}
\caption{Comparison between eikonal wavefront computed numerically and experimental data. The bright yellow dots represent the eikonal wavefront reconstructed from the phase along the rays. The two black circles are the unstable orbits, also present in Fig.~\ref{characteristics}. The background image is a measurement of the free surface of the water. The flow parameters and wave frequency are the same as in Fig.~\ref{Rays}. The frequency of the eikonal wave ($3.15$ Hz) is chosen so as to obtain the best fit with the $3.27$ Hz wave of Fig.~1 \cite{Superradiance}. The two frequencies agrees within error bars (about $4\%$). The colorbar represents the amplitude of the wave in metres.
}
\label{wavefront_exp}
\end{figure}

In Fig.~\ref{wavefront_exp} we compare our ray-tracing results with experimental data taken from the wavetank  experiment of \cite{Superradiance} with a circulation-to-draining ratio of approximately $C/D \simeq 16$. One can see that, broadly, the eikonal wavefronts agree well with the experiment. We observe small deviations in two regions: close to the centre of the vortex, and after the wave has propagated through the vortex (on the left of the image). The eikonal approximation we used is expected to degrade close to the centre of the vortex, where the flow varies more rapidly. Moreover, cumulative errors might explain the shift between our simulation and the data after the wave has propagated away from the vortex (on the far left side of Fig.~\ref{wavefront_exp}). 

The good agreement between our numerical solution and the experimental data suggests that circular rays are present in the system (see Sec.~\ref{subsec:circorbits}). While the inner one (co-rotating with the vortex) is very close to the centre, in a region where the vortex model of \eqref{DBT} becomes questionable, the outer one (counter-rotating) lies in a region where our method provides an accurate description. To estimate the order of the error, we compare the gradient scale of the background to the frequency of the wave: $|\partial_r v_\theta| / \omega$. Near the outer light ring $r_\star^+$, this ratio reduces to the inverse of the azimuthal number $m_\star$ and is approximately $0.1$. As a last remark, we point out that within our approximation, the location of the rays are usually more accurate than the phase itself. This is because the phase is rapidly varying, but the local momentum is slowly varying (see the discussion in section 4.3.4 of \cite{Buhler}). Hence, this fact and the good agreement in figure~\ref{wavefront_exp} strengthens our conclusion regarding the presence of circular orbits around the vortex of Fig.~\ref{wavefront_exp}. 

\section{Characteristic damped resonances of a draining vortex: the quasi-normal mode spectrum}
\label{QNM_Sec}
We anticipate that the presence of `rings' (i.e.~circular rays/characteristics) in the vortex system will influence the time-dependent response to a small wave excitation, through certain characteristic damped oscillations. In the black hole case, the time-dependent response of a black hole to a perturbation typically bears the imprint of one or more  Quasi-Normal Modes (QNM)~\cite{Berti:2009kk}. The QNMs are a discrete set of modes with \emph{complex} frequencies, with the real part determining the oscillation frequency, and the (negative) imaginary part determining the damping rate. Formally, QNMs are defined via a pair of boundary conditions, to be those modes which are ingoing at the horizon and outgoing far from the black hole. In the eikonal limit, one finds that the spectrum is governed chiefly by the properties of the circular orbits (see e.g.~\cite{Cardoso}). More precisely, the eikonal complex frequencies are given by 
\be \label{QNM_eiko}
\omega_{\rm QNM}(m) = \omega_\star(m) - i \Lambda(m) \left(n+\frac{1}{2} \right) , 
\ee
where $\omega_\star(m)$ is the circular orbit wave frequency of Sec.~\ref{subsec:circorbits}, $n$ is an integer called the overtone index, and $\Lambda$ is the decay rate of these (unstable) orbits (Lyapunov exponent), to be defined below.

The notion of QNMs extends to many different contexts. QNMs typically arise in open systems, or systems with leakage (in our case, the drain plays that role). As we shall see, around a draining vortex, surface wave decay is also driven by a discrete set of complex frequencies, that can be obtained approximately using the structure of circular orbits. 

Due to the presence of circular orbits, we expect that, if one perturbs the system initially, the perturbations will spread out rapidly, except where the perturbation lingers around the circular orbit. Indeed, a wave packet passing close to such a `ring' can hover around it for a longer time before dispersing. To understand this behaviour more precisely, we consider a wave packet in the neighbourhood of a circular orbit, that is, around the family of rays $(r_\star + \delta r, k_{r \star} + \delta k, \omega_\star + \delta \omega)$ at a fixed azimuthal number $m$. To obtain the effective dynamics of these rays, we expand the Hamiltonian in the vicinity of the critical point $(r_\star, k_{r \star})$. This gives  
\be \label{LR_Ham}
\mathcal{H} = \partial_{\omega}\mathcal{H} \delta \omega + \frac{1}{2} X^{T} \cdot \left[ d^{2}\mathcal{H} \right] \cdot  X, 
\ee
where we have defined the Hessian matrix 
\be
[d^{2}\mathcal{H}] =\begin{pmatrix}
\partial_{k}^{2}\mathcal{H} & \partial_{k}\partial_r \mathcal{H} \\
\partial_{k}\partial_{r} \mathcal{H} & \partial_{r}^{2}\mathcal{H} 
\end{pmatrix} \qquad \text{and} \qquad 
X = \begin{pmatrix} \delta k \\ \delta r \end{pmatrix}.
\ee
Since $[d^{2}\mathcal{H}]$ is a symmetric and real matrix, it can be diagonalized. We call $(\mu_{1},\mu_{2})$ the eigen-values, and $(R,K)$ the components in the eigen-basis\footnote{In addition, the eigen-basis is orthogonal, which in two dimensions is enough to ensure a canonical transformation $(\delta r, \delta k) \to (R,K)$.}. In this representation, the Hamiltonian becomes
\be \label{Red_H}
\mathcal{H} = \partial_{\omega}\mathcal{H} \delta \omega + \frac{1}{2}\mu_{1}K^{2} + \frac{1}{2}\mu_{2} R^{2}.
\end{equation}
This is the Hamiltonian of a harmonic oscillator. The relative sign of the coefficients $\mu_1$ and $\mu_2$ will determine the stability of the orbits. In our case, we always have $\mu_{1}\mu_{2} < 0$ and hence the orbit is unstable. In its vicinity, wave dynamics is described by an \emph{inverted} harmonic oscillator. To obtain the Lyapunov exponent $\Lambda$, we solve Hamilton's equations with this reduced Hamiltonian. This gives  
\be \label{EoM_IHO}
\dot{R}=\mu_{1}K, \qquad \dot{K}=-\mu_{2}R, \qquad \dot{t}=-\partial_{\omega}\mathcal{H}.
\ee
The general solution is a linear superposition of decaying and growing exponentials. At late times, the growing behavior dominates, and we have
\begin{equation}
R \sim A e^{ \frac{\sqrt{-\mu_{1}\mu_{2}}}{\partial_{\omega}\mathcal{H}}t} \qquad \text{and} \qquad K \sim B e^{ \frac{\sqrt{-\mu_{1}\mu_{2}}}{\partial_{\omega}\mathcal{H}}t},
\end{equation}
where $A$ and $B$ are constant coefficients such that \eqref{EoM_IHO} is satisfied. The Lyapunov exponent is the rate of exponential growth away from the orbit, and is given by 
\begin{equation}
\Lambda = \frac{\sqrt{-\mu_{1}\mu_{2}}}{|\partial_{\omega}\mathcal{H}|}=\frac{\sqrt{-\det[d^{2}\mathcal{H}]}}{|\partial_{\omega}\mathcal{H}|}.
\end{equation}
The Lyapunov exponent governs the decay of waves close to the circular orbit. One can see this by inspecting the amplitude governed by the transport equation \eqref{transport_eq}, namely,
\be
A_0(t) \propto e^{-\Lambda t/2}. 
\ee
However this argument misses the overtone frequencies (i.e. $n \neq 0$ in \eqref{QNM_eiko}). 

A more precise argument is to lift the reduced Hamiltonian \eqref{Red_H} at the level of the wave equation. To see this, we consider a wave packet of fixed $m$, localized around the circular orbit, written as 
\be \label{LR_SVA}
\phi(t,r_\star +\delta r) \sim \psi(t,\delta r) e^{- i \omega_\star t + i k_{r \star} r}, 
\ee
where $\psi$ is a slowly-varying envelope. Formally, the wave equation \eqref{wave_eq} can be written\footnote{This writing is of course only formal, since there is \emph{a priori} no unique way to promote the Hamiltonian function $\mathcal H(\omega, r, k_r)$ into an operator $\hat{\mathcal{H}}(i \partial_t; r,-i\partial_r)$. The ambiguities essentially comes from the various possible ordering between functions of $r$ and $\partial_r$. However, since we work here at the level of the eikonal approximation, different choices of ordering lead to the same result.}
\be
\hat{\mathcal H}(\omega= i \partial_t; r, k_r = -i\partial_r) \phi = 0. 
\ee
Assuming $\psi$ in \eqref{LR_SVA} is slowly varying, one can again rescale $\partial \to \epsilon \partial$ and expand in $\epsilon$. Doing so, the effective wave equation in the vicinity of the circular orbit is given by
\be \label{Brut_Schro_eq}
- i\partial_{\omega}\mathcal{H} \partial_t \psi = -\frac{1}{2} \partial_k^2 \mathcal H \partial_r^2 \psi - \frac{i}{2} \partial_k \partial_r \mathcal H (r \partial_r + \partial_r r) \psi + \frac{1}{2} \partial_r^2 \mathcal H r^{2} \psi. 
\ee
(Note that the cross term is chosen to be symmetric.) This equation is a Schr\"odinger equation in an upside-down harmonic potential. To see this, we use the fact that there is a canonical transformation to diagonalize the Hamiltonian \eqref{LR_Ham}. At the level of the Schr\"odinger equation \eqref{Brut_Schro_eq}, this means that there is a unitary transformation $\psi \to \tilde \psi$ such that 
\be \label{Schro_eq}
- i\partial_{\omega}\mathcal{H} \partial_t \tilde \psi = -\frac{1}{2}\mu_{1} \partial_R^2 \tilde \psi + \frac{1}{2}\mu_{2} R^{2} \tilde \psi . 
\ee
In the case of Eq.~\eqref{Schro_eq} the time dependent response is well described with the help of quasi-normal modes, defined as the solutions of $i \partial_t \tilde \psi = \delta \omega \; \tilde \psi$ with purely out-going boundary conditions (see e.g. \cite{Kokkotas99}). Such solutions are analytically known, and can be expressed using parametric cylinder functions. This selects a discrete set of complex values for $\delta \omega$, which are the quasi-normal frequencies for $\tilde \psi$ and hence for $\psi$. Finally, using Eq.~\eqref{LR_SVA}, we see that the quasi-normal frequencies for $\phi$ are given by Eq.~\eqref{QNM_eiko}.

Alternative boundary conditions, such as those for exotic black hole mimickers, can alter the global structure of the quasinormal mode spectrum, without significantly altering the local `ringing' phenomena associated with the light-ring (see e.g.~\cite{Cardoso:2017cqb, Cardoso:2017njb}).

\begin{figure}
\begin{subfigure}{.5\textwidth}
\centering
  \includegraphics[width=1\linewidth]{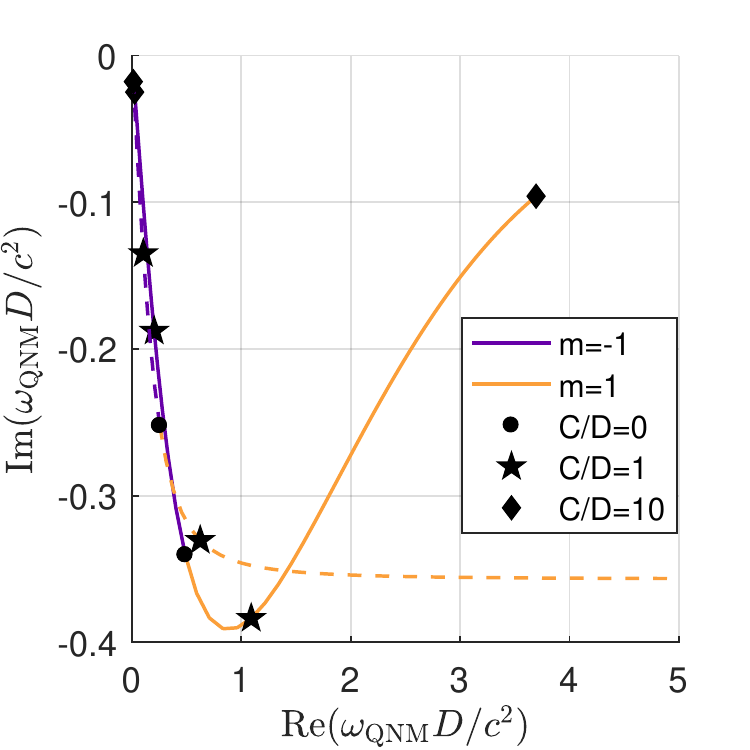}
  \caption{$m = \pm 1$}
\end{subfigure}%
\begin{subfigure}{.5\textwidth}
  \includegraphics[width=1\linewidth]{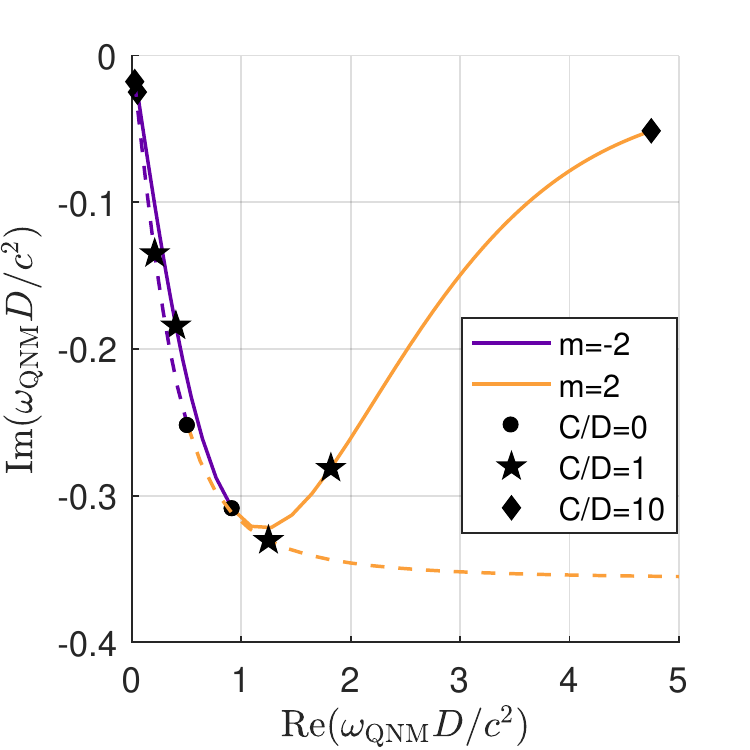}
  \caption{$m = \pm 2$}
\end{subfigure}
\caption{Fundamental quasi-normal mode frequencies $\omega_{\rm QNM}$ of different $m$ modes, $m = \pm 1$ (left) and $m = \pm 2$ (right), for various ratio $C/D$. The dashed line is the shallow water result, and the solid one the (general) dispersive regime. In order to make the transition more transparent, the depth is chosen so that $g h^{3}/D^2 = 0.04$. Each point represents a different ratio for the flow parameters. The negative $m$ modes have a similar behaviour in the two regimes, namely their lifetime increases as the ratio $C/D$ increases. On the other hand, the lifetime of the positive $m$ modes increases too with the ratio $C/D$ in the dispersive regime while it monotonically decrease in the absence of dispersion. 
}
\label{QNM_plot}
\end{figure}

We have numerically computed the fundamental  ($n = 0$) eikonal quasi-normal frequencies in the full dispersive regime for a selection of azimuthal numbers and for several values of the ratio $C/D$. The result, and the comparison with the linear regime is presented in Fig.~\ref{QNM_plot}. We can see that the behaviour of the counter-rotating modes ($m<0$) is very similar in the two different regimes. As the ratio increases, the radius of the unstable orbits of these modes will increase too. For fixed $m$, the increase of the radius will increase the wavelength (decrease the frequency) and therefore will bring the mode closer to the linear regime. On the other side, the behaviour for co-rotating ($m>0$) modes is very different when dispersion is included. Indeed, in the relativistic regime, the lifetime of the co-rotating modes decreases while it appears to increase after a short drop as an effect of dispersion.

In general, it is noticeable that the co-rotating modes are damped quicker than the counter-rotating ones (see Fig.~\ref{QNM_plot}). Hence, if the initial excitation has a similar amplitude on both radius, the dominant signal at large time will be that of the counter-rotating orbit. This is the case for instance if the system is excited by sending a plane wave, which contains as much $m>0$ as $m<0$ (as in Fig.~\ref{wavefront_exp}). In addition, we expect the counter-rotating mode to have a larger spatial extension than the co-rotating one since the radius of the counter-rotating orbit is larger that the co-rotating one, and hence, be more visible. 

\section{Discussion}

It is known that waves in a shallow potential flow are governed, approximately, by a wave equation equivalent to that governing a scalar field on an effective spacetime. By means of this analogy, peculiar and fundamental effects which were before out of reach, are now observable in the laboratory. Recent water-wave experiments with draining, circulating flows have found that one of these analogue phenomena, namely black hole superradiance, persist beyond the shallow regime, where dispersive and dissipative effects render incomplete the spacetime description of the system. In this work, we have applied a ray-tracing (gradient expansion) method to a realistic dispersive, dissipative wave equation. We argued that fundamental features of vortex wave scattering can be understood with reference to a pair of (frequency-dependent) circular rays (rings), which are analogous to the light rings of black holes. 

In addition, we have reconstructed the wavefronts of a plane wave scattering on a vortex using the ray-tracing approximation. This allows us to test our methods against the experimental results of \cite{Superradiance} (see Fig.~\ref{wavefront_exp}). The agreement with the data suggests the presence of an outer light ring (for counter-rotating modes) and raises the challenge of observing it. This could be done through the observation of quasi-normal modes of the vortex flow that are a generic consequence of light rings. We discussed these quasi-normal modes and their relation to circular orbits in section \ref{QNM_Sec}. In particular we have shown that the effect of dispersion on the counter rotating modes will be the most important when the drain predominates over the circulation. This is the case in the experiment of \cite{Superradiance}, where the ratio $C/D \simeq 16$. Another interesting feature of dispersion is that the lifetime of the QNM increases at high circulation. For the co-rotating mode, this behaviour significantly differs from the non-dispersive (shallow water) regime and might lead to an instability if the decay of those modes becomes smaller than their amplification via superradiance (\cite{OLI14, HOD14,BRI15}). As a last remark, we would like to point out that these phenomena may also be relevant on oceanic scales, for instance \cite{Haller13} have shown the presence of similar analogue black hole light rings in the South Atlantic portion of the Atlantic Ocean.

\section*{Acknowledgement}
We want to thank S. Patrick and Z. Fifer for the many useful discussions at various stages of the project. We also thank T. Napier and T. Wright for their help and guidance with the experimental set up.  S.D. thanks David Dempsey for helpful discussions.
This project has received funding from the European Union's Horizon 2020 research and innovation programme under the Marie Sk\l odowska-Curie grant agreement No 655524. 
SW acknowledges financial
support provided under the Royal Society University Research Fellow (UF120112), the
Nottingham Advanced Research Fellow (A2RHS2), the Royal Society Project (RG130377)
grants, and the EPSRC Project Grant (EP/P00637X/1). SW acknowledge partial support from STFC consolidated grant No. ST/P000703/.
S.D. acknowledges financial support from the Engineering and Physical Sciences Research Council (EPSRC) under Grant No. EP/M025802/1, from the Science and Technology Facilities Council (STFC) under Grant No. ST/L000520/1 and the project H2020-MSCA-RISE-2017 Grant FunFiCO-777740.

\appendix

\section{General method of ray tracing}
\label{WKB_App}
In this appendix we review the general framework for solving Eq.~\eqref{wave_eq} using a gradient expansion (see for example~\cite{Buhler}). To do so, we consider the rescaling of Eq.~\eqref{wave_eq} as follows: 
\bsub \bea
\partial &\to& \epsilon \partial , \\
\nu &\to& \epsilon \nu. 
\eea \esub
The second line translates the fact that we consider dissipation to be weak (over a scale of the order of the gradient scale) (it is also more natural to impose a real-valued Hamilton-Jacobi equation~\eqref{H-J}, which imposes to include dissipation in the amplitude equation). As in the text, we assume a solution to the wave equation of the form 
\be 
\phi = A \exp \left(i\frac S \epsilon\right). 
\ee
where $\phi$, $A$ and $S$ are functions of $\vec{x}$ and $t$. $A$ and $S$ can be expanded as power series in the small parameter $\epsilon$: 
\begin{equation} \label{expansion}
S=\sum_{n=0}^{+\infty} \epsilon^n S_{n} \qquad \text{and} \qquad A= \sum_{n=0}^{+\infty} \epsilon^n A_{n} .
\end{equation}
The role of $\epsilon$ is to keep track of the hierarchy of terms in the gradient expansion. Intuitively, it can be seen as the ratio between the gradient scale and the wavelength. 
We now seek the leading-order equations for both $S$ and $A$. The obtained solutions gives us approximate wave solutions using \eqref{eikonal_wave} that we shall refer to as \textit{eikonal waves}.

\subsection{Ray tracing equations}

We start from the wave equation \eqref{wave_eq} with the \emph{ansatz} \eqref{eikonal_wave} for $\phi$ and keep only the leading order term in $\epsilon$. To do so we expand the function $F$ as a Taylor expansion :
\begin{equation}
F(-i\epsilon \bnab )= \sum_{n=1}^{+\infty} T_{2n} (-1)^n \epsilon^{2n} \Delta^{n} , \label{F_series}
\end{equation}
where $\Delta \equiv \nabla^2 = \partial_x^2 + \partial_y^2$ is the 2D Laplacian. To expand the term $F(-i\bnab ) \phi$ of Eq.~\eqref{wave_eq}, we look at each term of the Taylor series \eqref{F_series} separately. Using the binomial formula, we obtain 
\bea
\Delta^{n} \phi &=& \sum_{k=0}^{n} \binom{n}{k} \partial_{x}^{2k}\partial_{y}^{2n - 2k} \phi , \label{Delta_expansion} \\
&\simeq& \frac{(-1)^n}{\epsilon^{2n}} \sum_{k=0}^{n} \binom{n}{k} (\partial_x S_0)^{2k} (\partial_{y} S_0)^{2n - 2k} A e^{i S_0/\epsilon} , \nonumber \\
&\simeq& \frac{(-1)^n}{\epsilon^{2n}} \left((\partial_x S_0)^2 + (\partial_y S_0)^2\right)^n A e^{i S_0/\epsilon} , \nonumber
\eea 
where $\simeq$ means that we have kept only the leading term in $\epsilon$. Re-summing the Taylor series \eqref{F_series}, we see that 
\be \label{action_of_F}
F(-i\epsilon \bnab ) \phi \simeq A e^{i S_0/\epsilon} F(\bnab S_0). 
\ee
The same applies for the material derivative, that is 
\be \label{action_of_D}
\mathcal{D}_t^{2}\phi \simeq -A e^{iS/\epsilon}(\mathcal{D}_t S_{0})^{2}. 
\ee
Combining \eqref{action_of_F} and \eqref{action_of_D}, it follows that at leading order, the phase $S_0$ obeys the Hamilton-Jacobi equation \eqref{H-J}.

\subsection{Transport equation}
Above, we derived the ray equations from the leading-order-in-$\epsilon$ term of the wave equation. We now focus on the next-to-leading order, and show that it leads to a transport equation for the amplitude of the wave. Let us first examine the next-to-leading-order term of the action of $\mathcal{D}_t^{2}$ on $\phi$. As $\mathcal{D}_t^{2}$ contains two derivatives, the next-to-leading-order term of its action on $\phi$ is attained when only one derivative hits the exponential. 
\begin{equation}
\mathcal{D}_t^{2} \phi \simeq  \frac{-i \epsilon}{A_0} \mathcal{D}_t\left(A_{0}^2 \Omega_0 \right)
\end{equation}
where we have introduced $\Omega_{0}=\omega - \vec{v_0}.\bnab S_{0}$ the comoving angular frequency of the wave.
We get the action of the operator $F(-i\epsilon\bnab )$ the same way. 
To obtain the next to leading order term of $\Delta^n \phi$ (in $1/\epsilon^{2n-1}$), we must count the number of terms in the expansion of \eqref{Delta_expansion} such that the exponential $e^{iS/\epsilon}$ is derived exactly $2n-1$ times. The last derivative of $\Delta^n$ will then act on $A_0$ or $\bnab S_{0}$. Doing so, we obtain 
\begin{eqnarray}
\Delta^n \phi &\simeq& \frac{(-1)^{n-1}}{\epsilon^{2n-1}}i \bigg( 2n \bnab A_0 \cdot (\bnab S_{0})^{2n-1}  + \frac{2n(2n-1)}{2} A_0 \cdot (\bnab S_{0})^{2n-2} \Delta S_{0} \bigg) e^{iS/\epsilon}. \nonumber \\ && 
\end{eqnarray}
Re-summing the Tayor series, we recognise the derivative with respect to the momenta of the function $F$. Gathering the terms, the action of $F(-i\epsilon\bnab )$ can be rewritten as a total derivative: 
\begin{eqnarray}
F(-i\epsilon \bnab )A_{0}e^{iS/\epsilon} \simeq  -\frac{i\epsilon}{2 A_0} \bnab . \left( A_0^2 \nabla_k F(\nabla S_0) \right).
\end{eqnarray}
Finally, a similar calculation yields the action of the damping operator,
\be
2\epsilon \nu\gamma(-i\bnab)\mathcal D_t \phi \simeq 2i\epsilon \nu\Omega_0 A_0 \gamma(\bnab S_0).
\ee
Combining the action of the three operators we obtain the equation for the amplitude used in the text, i.e. Eq.~\eqref{transport_eq}.

\section{Orbits in the deep water regime} \label{App:orbits}
We detail here the derivation of Eq.~\eqref{rad_deep}. As we have seen, there are two different light rings (represented by the $\pm$ in \eqref{radius_rel}). In order to keep track of this sign we introduce $\epsilon_m = \text{sign}(m)$. Using it, we can express $m$ as a function of $r$ as 
\be
m = \epsilon_m \frac{g r_{\star}^3}{4 B_\pm^2}\sqrt{ 1 - \frac{D^2}{B_\pm^2}}. 
\ee
We shall also use the parameters $B_\pm$ introduced in Eq.~\eqref{B_param}. On a circular orbit, we can use the Hamiltonian constraint to express the orbital frequency in terms of the other parameters: 
\be
\omega_\star = -\frac{D}{r_\star} k_{r\star} + \frac{C}{r_\star} \frac{m}{r_\star} + \sqrt{g k}.
\ee
Substituting \eqref{p_deep} and \eqref{pr_m_deep} in the above equation, we obtain 
\be \label{om_star_app}
\omega_\star = \frac{gr_\star}{2B_\pm} \left( - \frac{D^2}{2B_\pm^2} + \epsilon_m \frac{C}{2B_\pm^2}\sqrt{B^2 - D^2} + 1 \right).
\ee
We now rewrite our flow parameters $C$ and $D$ as 
\be \label{new_CD}
C= R \sin(\phi) \quad \text{and} \quad D = R \cos(\phi),
\ee
and $B_\pm$ becomes 
\be \label{new_B}
B=\sqrt{2}R\sqrt{1 - \epsilon_m sin(\phi)}.
\ee
Inserting \eqref{new_CD} and \eqref{new_B} into \eqref{om_star_app} we get 
\be
\omega_\star = \frac{gr_\star}{8B_\pm} \left( - \frac{\cos^2(\phi)}{1 - \epsilon_m \sin(\phi)} + \epsilon_m \frac{\sin(\phi) \sqrt{2 - 2\epsilon_m \sin(\phi) - \cos^2(\phi)}}{ 1 - \epsilon_m \sin(\phi)} 
+ 4 \right). \nonumber
\ee
Using $\cos^2(\phi) = 1 - \sin^2(\phi) $, we can simplify this expression:
\be
\omega_\star = \frac{gr_\star}{8B_\pm} \left( \frac{ -1 + \sin^2(\phi) }{1 - \epsilon_m\sin(\phi)} + \epsilon_m \sin(\phi)  + 4 \right), 
\ee
which further simplifies into 
\be
\omega_\star = \frac{3gr_\star}{8B_\pm}.
\ee
We can inverse this relation to obtain $r_\star$ as a function of $\omega_\star$ or express $r_\star$ in terms of $m$ to obtain the effective dispersion relation \eqref{effective_dr}.

\section{Velocity profiles} \label{App:velocities}

Here we present the velocity profiles used in our model in order to compare the various quantities entering in our discussion. 

%\begin{figure}
%\centering
 % \includegraphics[width=0.65\linewidth]{Figs/velocity_profiles.pdf}
%\caption{Radial (in red) and angular (in blue) velocity profiles as functions of the radius r. The profiles are plotted for \mbox{$C=1.6\times 10^{-2} m^2s^{-1}$} and \mbox{$D=1\times 10^{-3} m^2s^{-1}$}. The dashed vertical black lines represent the values of the two circular orbits. The dotted horizontal black line depicts the value of the group velocity and phase for the frequency considered in figure 3 (3.15Hz). As the radial velocity is much smaller than the angular one, the total velocity is approximately equal to the angular one and it matches the group velocity of the wave at about 7cm.}
%\end{figure}

\begin{center}
\includegraphics[width=0.8\linewidth]{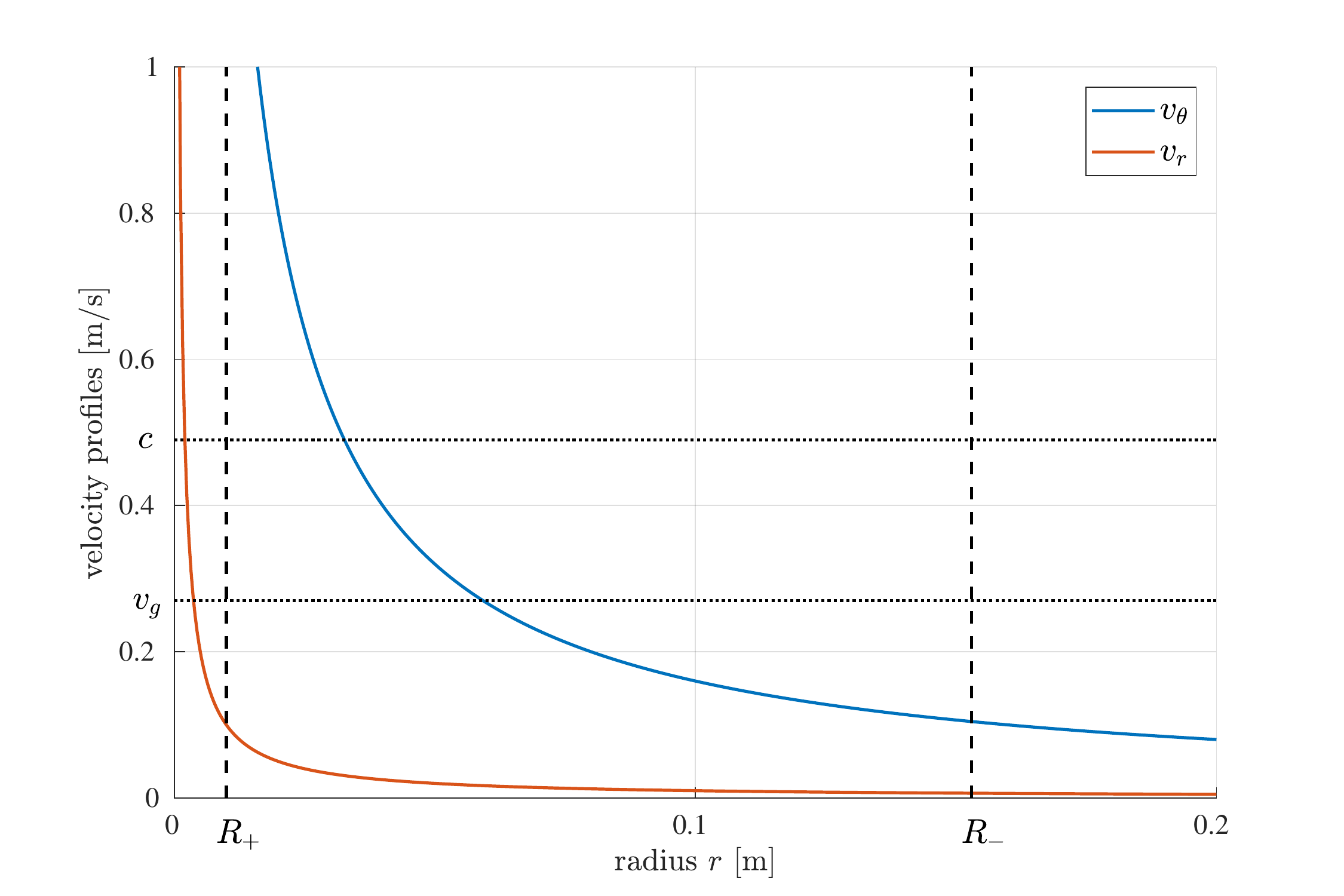}
\end{center}
{\noindent \textbf{Figure 6:} Radial (in red) and angular (in blue) velocity profiles as functions of the radius $r$. The profiles are plotted for \mbox{$C=1.6\times 10^{-2} m^2s^{-1}$} and \mbox{$D=1\times 10^{-3} m^2s^{-1}$}. The dashed vertical black lines represent the values of the two circular orbit radii. The dotted horizontal black lines depict the value of the group velocity and phase velocity of the wave for the frequency considered in Fig. \ref{Rays} (3.15Hz). As the radial velocity is much smaller than the angular one, the total velocity is approximately equal to the angular one and it matches the group velocity of the waves at about 7cm.}

\newpage
\bibliographystyle{jfm}
\bibliography{bibli_geo}

\end{document}